\documentclass{sig-alternate}

\usepackage{listings}
\usepackage{graphicx}
\usepackage[linesnumbered, ruled]{algorithm2e}
\usepackage{enumitem}
\usepackage{url}

\begin{document}

\conferenceinfo{WHPCF}{'13 Denver, Colorado USA}
\CopyrightYear{2013}

\title{Accounting for Secondary Uncertainty:\\Efficient Computation of Portfolio Risk Measures\\on Multi and Many Core Architectures}

\numberofauthors{1}
\author{
\alignauthor
Blesson Varghese and Andrew Rau-Chaplin\\
       \affaddr{Faculty of Computer Science, Dalhousie University, Halifax, Nova Scotia, Canada}\\
       \email{\{varghese, arc\}@cs.dal.ca}
}

\date{25 August 2013}

\maketitle

\begin{abstract}
Aggregate Risk Analysis is a computationally intensive and a data intensive problem, thereby making the application of high-performance computing techniques interesting. In this paper, the design and implementation of a parallel Aggregate Risk Analysis algorithm on multi-core CPU and many-core GPU platforms are explored. The efficient computation of key risk measures, including Probable Maximum Loss (PML) and the Tail Value-at-Risk (TVaR) in the presence of both primary and secondary uncertainty for a portfolio of property catastrophe insurance treaties is considered. {\it Primary Uncertainty} is the the uncertainty associated with whether a catastrophe event occurs or not in a simulated year, while {\it Secondary Uncertainty} is the uncertainty in the amount of loss when the event occurs.

A number of statistical algorithms are investigated for computing secondary uncertainty. Numerous challenges such as loading large data onto hardware with limited memory and organising it are addressed. The results obtained from experimental studies are encouraging. Consider for example, an aggregate risk analysis involving 800,000 trials, with 1,000 catastrophic events per trial, a million locations, and a complex contract structure taking into account secondary uncertainty. The analysis can be performed in just 41 seconds on a GPU, that is 24x faster than the sequential counterpart on a fast multi-core CPU. The results indicate that GPUs can be used to efficiently accelerate aggregate risk analysis even in the presence of secondary uncertainty.
\end{abstract}





\section{Introduction}
\label{introduction}

Reinsurance companies who insure primary insurance companies against losses caused by catastrophes, such as earthquakes, hurricanes and floods, must quantify the risk related to large portfolios of risk transfer treaties. In reinsurance a portfolio represents complex insurance contracts covering properties against losses due catastrophes, and contracts include Per-Occurrence eXcess of Loss (XL), Catastrophe XL and Aggregate XL treaties. Aggregate Risk Analysis is performed on portfolios to compute risk measures including Probable Maximum Loss (PML) \cite{PML-1} and the Tail Value-at-Risk (TVaR) \cite{TVAR-2}. Such an analysis is central to treaty pricing and portfolio/solvency applications. The analysis may involve simulations of over one million trials in which each trial consists of one thousand catastrophic events each of which may impact tens of thousands to millions of individual properties, for example buildings. Not only is the analysis computationally intensive but also data intensive, and therefore the application of High Performance Computing (HPC) techniques is desirable.

The analysis can be run weekly, monthly or quarterly on production systems based on the requirement for updating the portfolio. For example, based on the fluctuation of currency rates an entire portfolio a weekly update can be performed which requires more than twenty four hours. Often times it is not sufficient to run routine analysis but requires ad hoc analysis. For example, consider a real-time pricing scenario in which an underwriter can evaluate different contractual terms and pricing while discussing with a client over a telephone. This cannot be accommodated on production system that is committed to routine analysis and the response required for real-time cannot be achieved on these systems. Hence, achieving significant speed up using high-performance computing techniques in risk analysis is desirable.

In our previous work \cite{SC2012}, we explored the design and implementation of a parallel Aggregate Risk Analysis algorithm which was significantly faster than previous sequential solutions. However, it was limited for its use in portfolio wide risk analysis scenarios since the algorithm could only account for {\it Primary Uncertainty} - the uncertainty whether a catastrophic event occurs or not in a simulated year. It was not able to account for {\it Secondary Uncertainty}, the uncertainty in the amount of loss incurred when the event occurs. In this paper, secondary uncertainty is taken into account whereby loss distributions flow through the simulation rather than mean loss values due to an event.

In practice there are many sources of secondary uncertainty in catastrophic risk modelling. For example, the exposure data which describes the buildings, their locations, and construction types may be incomplete, lacking sufficient detail, or may just be inaccurate. Also physical modelling of hazard, for example an earthquake, may naturally generate a distribution of hazard intensity values due to uncertainty in energy attenuation functions used or in driving data such as soil type. Lastly, building vulnerability functions are simplifications of complex physical phenomenon and are therefore much better at producing loss distributions than accurate point estimates. An aggregate risk analysis algorithm that accounts only for primary uncertainty uses only mean loss values and fails to account for what is known about the loss distribution. Therefore, aggregate analysis needs to take both primary and secondary uncertainty into account by considering event loss distributions represented by the event occurrence probability, mean loss, and independent and correlated standard deviations. This captures a wide range of possible outcomes.

The research reported in this paper proposes an aggregate risk analysis algorithm capable of capturing both primary and secondary uncertainty. The algorithm is designed to run efficiently on both multi-core CPUs and many-core GPUs. A distribution of losses is used in the simulation rather than just mean loss values and efficient statistical operations, such as Cumulative Distribution Functions and Quantiles of the Normal and Beta distributions need to be performed. A significant challenge to not only balance the workload across threads performing fixed time operations (for example, addition operation), but also to balance the workload when individual numerical operations require variable time (for example, computations in iterative methods) for achieving effective parallelism is addressed. The implementation and optimization of the algorithm for multi and many core architectures is presented along with experimental evaluation. Since performance of the algorithm is dependent on the underlying statistical operations, four different statistical libraries were explored for use on the GPU and an additional three statistical libraries for use on the CPU. A parallel simulation of 800,000 trials with 1,000 catastrophic events per trial on an exposure set and on a contract structure taking secondary uncertainty exhibits a speedup of 24x over the sequential implementation on the CPU.

The remainder of this paper is organized as follows. Section \ref{finegrainanalytics} proposes an algorithmic framework for performing Aggregate Risk Analysis with Primary and Secondary Uncertainty. Section \ref{secondaryuncertainty} presents how secondary uncertainty is applied within the inner loop of the risk analysis algorithm. Section \ref{implementation} describes the implementation of the proposed algorithm on the GPU platform. Section \ref{results} highlights the results obtained from experimental evaluation. Section \ref{conclusion} concludes the paper by considering future work.

\section{Aggregate Risk Analysis with Primary and Secondary Uncertainty}
\label{finegrainanalytics}

Stochastic Monte Carlo simulations are required for portfolio risk management and contract pricing. Such a simulation in which each trial of the simulation represents a distinct view of which catastrophic events occur and in what order they occur in a contractual year is referred to as Aggregate Risk Analysis \cite{s2, s3, s5}. One merit of performing such an analysis is that millions of alternative views of a single contractual year can be obtained. This section considers the inputs required for performing aggregate risk analysis, proposes an algorithm for aggregate risk analysis, considers the financial terms employed in the algorithms, and presents the output of the analysis.

\subsection{Inputs}
Three data tables are input to Aggregate Risk Analysis. The first table is the Year Event Table (YET), denoted as $YET$, which is a database of pre-simulated occurrences of catastrophic events from a catalogue of stochastic events. Each record in a YET called a ``trial'', denoted as $T_i$, represents a possible sequence of event occurrences for any given year. The sequence of events is defined by an ordered set of tuples containing the ID of an event and the time-stamp of its occurrence in that trial
\begin{equation*}
\begin{array}{l c l}
T_i	& =	& \{(E_{i, 1}, t_{i, 1}, z_{(Prog, E)_{i, 1}}), \dots,\\
	&	& (E_{i, k}, t_{i, k}, z_{(Prog, E)_{i, k}})\}.
\end{array}
\end{equation*}

The set is ordered by ascending time-stamp values. Program-and-Event-Occurrence-Specific random number, $z_{(Prog, E)}$ is considered in Section \ref{secondaryuncertainty}. A typical YET may comprise thousands to a million trials, and each trial may have approximately between 800 to 1500 `event time-stamp' pairs, based on a global event catalogue covering multiple perils. The YET can be represented as
\begin{equation*}
\begin{array}{{l c l}}
YET	&	=	& \{ T_i = \{(E_{i, 1}, t_{i, 1}, z_{(Prog, E)_{i, 1}}), \dots,\\
	&		& (E_{i, k}, t_{i, k}, z_{(Prog, E)_{i, k}})\} \},
\end{array}
\end{equation*}
\begin{center}
where $i = 1, 2, \dots$ and $k = 1, 2, \dots, 800-1500$.
\end{center}

The second table is the Extended Event Loss Tables, denoted as $XELT$, which represents a collection of specific events and their corresponding losses with respect to an exposure set. In addition, a few parameters, namely the Event-Occurrence-Specific random number ($z_{(E)}$), the independent standard deviation of loss ($\sigma_{I}$), the correlated standard deviation of loss ($\sigma_{C}$), and the  maximum expected loss ($max_{l}$) are represented within the $XELT$. The loss associated with an event $E_{i}$ is represented as $\mu_{l}$ is required for the analysis with secondary uncertainty. Applying secondary uncertainty using the XELT is presented in Section \ref{secondaryuncertainty}.

Each record in an XELT is denoted as `eXtended' event loss
\begin{equation*}
\begin{array}{l c l}
XEL_{i} &	=	& \{E_{i}, l_{i}, z_{(E)_{i}}, \sigma_{I_{i}}, \sigma_{C_{i}}, max_{l_{i}}\}.
\end{array}
\end{equation*}
and the financial terms associated with the XELT are represented as a tuple
\begin{equation*}
\begin{array}{l c l}
\mathcal{I} &	=	& (\mathcal{I}_{1}, \mathcal{I}_{2}, \dots).
\end{array}
\end{equation*}

A typical aggregate analysis may comprise 10,000 XELTs, each containing 10,000-30,000 extended event losses with exceptions even up to 2,000,000 extended event losses. The XELTs can be represented as
\begin{equation*}
XELT=\left\{
	\begin{array}{l c l}
	XEL_{i}		&	=	&	\{E_{i}, \mu_{l_{i}}, z_{(E)_{i}},\\
				&		& 	\sigma_{I_{i}}, \sigma_{C_{i}}, max_{l_{i}} \},\\
	\mathcal{I} &	=	&	(\mathcal{I}_{1}, \mathcal{I}_{2}, \dots)
	\end{array}\right\}
\end{equation*}
\begin{center}
with $i = 1, 2, \dots , 10,000-30,000$.
\end{center}

The third table is the Portfolio, denoted as $PF$, which contains a group of Programs, denoted as $P$ and represented as
\begin{equation*}
\begin{array}{l c l}
PF & = & \{P_{1}, P_{2}, \cdots, P_{n}\}
\end{array}
\end{equation*}
\begin{center}
with $n = 1, 2, \dots, 10$.
\end{center}

Each Program in turn covers a set of Layers, denoted as $L$, which covers a collection of XELTs under a set of layer terms. A single layer $L_i$ is composed of two attributes. Firstly, the set of XELTs
\begin{equation*}
\begin{array}{l c l}
\mathcal{E} = \{XELT_1, XELT_2, \dots, XELT_j\},
\end{array}
\end{equation*}
and secondly, the Layer Terms, denoted as
\begin{equation*}
\begin{array}{l c l}
\mathcal{T} = (\mathcal{T}_{OccR}, \mathcal{T}_{OccL}, \mathcal{T}_{AggR}, \mathcal{T}_{AggL}).
\end{array}
\end{equation*}

A typical Layer covers approximately 3 to 30 individual XELTs. The Layer can be represented as
\begin{equation*}
L=\left\{
	\begin{array}{l c l}
	\mathcal{E}	& =	& \{XELT_1, XELT_2, \dots, XELT_j\}, \\
	\mathcal{T}	& = & (\mathcal{T}_{OccR}, \mathcal{T}_{OccL}, \mathcal{T}_{AggR}, \mathcal{T}_{AggL})
	\end{array}\right\}
\end{equation*}
\begin{center}
with $j = 1, 2, \dots, 3-30$.
\end{center}

\subsection{Algorithm}
The basic algorithm (line no. 1-17 shown in Algorithm \ref{algorithm1}) for aggregate analysis has two stages. In the first stage, data is loaded into local memory what is referred to as the preprocessing stage in this paper. In this stage $YET$, $XELT$ and $PF$, are loaded into memory.
\begin{algorithm}
\caption{Aggregate Risk Analysis with Primary and Secondary Uncertainty}
\label{algorithm1}
\SetAlgoLined
\DontPrintSemicolon

\SetKwInOut{Input}{Input}
\SetKwInOut{Output}{Output}

\BlankLine

\Input{$YET$, $XELT$, $PF$}
\Output{$YLT$}

\BlankLine

\For{each Program, $P$, in $PF$}{
	\For{each Layer, $L$, in $P$}{
		\For{each Trial, $T$, in $YET$}{
			\For{each Event, $E$, in $T$}{
				\For{each $XELT$ covered by $L$}{
					Lookup $E$ in the $XELT$ and find corresponding loss, $l_{E}$\;
					Apply Secondary Uncertainty to $l_{E}$\;
					Apply Financial Terms to $l_{E}$\;
					$l_{T} \leftarrow$ $l_{T}$ + $l_{E}$\;
				}
				Apply Occurrence Financial Terms to $l_{T}$\;	
				Apply Aggregate Financial Terms to $l_{T}$\;
			}
		}
	}
}			
Populate $YLT$ using $l_{T}$\;
\BlankLine
\end{algorithm}

In the second stage, the four step simulation executed for each Layer and for each trial in the YET is performed as shown below and the resulting Year Loss Table ($YLT$) is produced.

In the first step shown in line no. 6 in which each event of a trial and its corresponding event loss in the set of XELTs associated with the Layer are determined. In the second step shown in line nos. 7-9, secondary uncertainty is applied to each loss value of the Event-Loss pair extracted from an XELT. A set of contractual financial terms are then applied to the benefit of the layer. For this the losses for a specific event's net of financial terms $\mathcal{I}$ are accumulated across all XELTs into a single event loss shown in line no. 9. In the third step in line no. 11 the event loss for each event occurrence in the trial, combined across all XELTs associated with the layer, is subject to occurrence terms. In the fourth step in line no. 12 aggregate terms are applied. The next sub-section will consider how the financial terms are applied.

\subsection{Applying Financial Terms}
The financial terms applied on the loss values combined across all XELTs associated with the layer are Occurrence and Aggregate terms. Two occurrence terms, namely (i) Occurrence Retention, denoted as $\mathcal{T}_{OccR}$, which is the retention or deductible of the insured for an individual occurrence loss, and (ii) Occurrence Limit, denoted as $\mathcal{T}_{OccL}$, which is the limit or coverage the insurer will pay for occurrence losses in excess of the retention are applied. Occurrence terms are applicable to individual event occurrences independent of any other occurrences in the trial. The occurrence terms capture specific contractual properties of 'eXcess of Loss' treaties as they apply to individual event occurrences only. The event losses net of occurrence terms are then accumulated into a single aggregate loss for the given trial. The occurrence terms are applied as $l_{T} = min ( max ( l_{T} - \mathcal{T}_{OccR} ), \mathcal{T}_{OccL})$.

Two aggregate terms, namely (i) Aggregate Retention, denoted as $\mathcal{T}_{AggR}$, which is the retention or deductible of the insured for an annual cumulative loss, and (ii) Aggregate Limit, denoted as $\mathcal{T}_{AggL}$, which is the limit or coverage the insurer will pay for annual cumulative losses in excess of the aggregate retention are applied. Aggregate terms are applied to the trial's aggregate loss for a layer. Unlike occurrence terms, aggregate terms are applied to the cumulative sum of occurrence losses within a trial and thus the result depends on the sequence of prior events in the trial. This behaviour captures contractual properties as they apply to multiple event occurrences. The aggregate loss net of the aggregate terms is referred to as the trial loss or the year loss. The aggregate terms are applied as $l_{T} = min ( max ( l_{T} - \mathcal{T}_{AggR} ), \mathcal{T}_{AggL})$.

\subsection{Output}
The output of the algorithm for performing aggregate risk analysis with primary and secondary uncertainty is a loss value associated with each trial of the YET. A reinsurer can derive important portfolio risk metrics such as the Probable Maximum Loss (PML) and the Tail Value-at-Risk (TVaR) which are used for both internal risk management and reporting to regulators and rating agencies. Furthermore, these metrics flow into a final stage of the risk analytics pipeline, namely Enterprise Risk Management, where liability, asset, and other forms of risks are combined and correlated to generate an enterprise wide view of risk.

Additional functions can be used to generate reports that will aid actuaries and decision makers. For example, reports presenting Return Period Losses (RPL) by Line of Business (LOB), Class of Business (COB) or Type of Participation (TOP). Further, the output of the analysis can be used for estimating Region/Peril losses and for performing Multi-Marginal Analysis and Stochastic Exceedance Probability (STEP) Analysis.

\section{Applying Secondary Uncertainty}
\label{secondaryuncertainty}

The methodology to compute secondary uncertainty heavily draws on industry-wide practices. The inputs required for the secondary uncertainty method and the sequence of steps for applying uncertainty to estimate a loss are considered in this section.

\subsection{Inputs}

Six inputs are required for computing secondary uncertainty which are obtained from the Year Event Table (YET) and the `eXtended ELT' (XELT). The first input is $z_{(Prog,E)} = P_{(Prog,E)} \in U(0,1)$ referred to as the Program-and-Event-Occurrence-Specific random number. Each Event occurrence across different Programs have different random numbers. The second input is $z_{(E)} = P_{(E)} \in U(0,1)$ referred to as the Event-Occurrence-Specific random number. Each Event occurrence across different Programs have the same random number. The third input is $\mu_{l}$ referred to as the mean loss. The fourth input is $\sigma_{I}$ referred to as the independent standard deviation of loss and represents the variance within the event-loss distribution. The fifth input is $\sigma_{C}$ referred to as the correlated standard deviation of loss and represents the error of the event-occurrence dependencies. The sixth input is $max_{l}$ referred to as the maximum expected loss.

\subsection{Combining standard deviation}
Given the above inputs, the independent and correlated standard deviations need to be combined to reduce the error in estimating the loss value associated with an event. This is done in a sequence of five steps. In the first step, the raw standard deviation is produced as $\sigma = \sigma_{I} + \sigma_{C}$. 

In the second step, the probabilities of occurrences, $z_{(Prog,E)}$ and $z_{(E)}$ are transformed from uniform distribution to normal distribution using
\begin{eqnarray*}
f(x; \mu, \sigma^{2}) = \int\limits_{-\infty}^{x} \frac{1}{\sigma\sqrt{2\pi}}e^{-\frac{1}{2}\big({\frac{x - \mu}{\sigma}}\big)^{2}}\mathrm{d}x
\end{eqnarray*}
This is applied to the probabilities of event occurrences as
\begin{eqnarray*}
\begin{array}{l c l}
v_{(Prog,E)} 	& = 	& f(z_{(Prog,E)}; 0, 1) \in N(0,1)\\
v_{(E)} 		& = 	& f(z_{(E)}; 0, 1) \in N(0,1)
\end{array}
\end{eqnarray*}

In the third step, the linear combination of the transformed probabilities of event occurrences and the standard deviations is computed as
\begin{eqnarray*}
LC = v_{(Prog,E)}\Big(\frac{\sigma_{I}}{\sigma}\Big) + v_{(E)}\Big(\frac{\sigma_{C}}{\sigma}\Big)
\end{eqnarray*}

In the fourth step, the normal random variable is computed as
\begin{eqnarray*}
v = \frac{LC}{\sqrt{{\big(\frac{\sigma_{I}}{\sigma}\big)}^2 + {\big(\frac{\sigma_{C}}{\sigma}\big)}^2}}
\end{eqnarray*}

In the fifth step, the normal random variable is transformed from normal distribution to uniform distribution as
\begin{eqnarray*}
z = \Phi(v) = F_{Norm}(v) = \frac{1}{\sqrt{2\pi}}\int\limits_{-\infty}^{v}e^{\frac{-t^2}{2}}\mathrm{d}t
\end{eqnarray*}

The model used above for combining the independent and correlated standard deviations represents two extreme cases. The first case in which $\sigma_{I} = 0$ and the second case in which $\sigma_{C} = 0$. The model also ensures that the final random number, $z$, is based on both the independent and correlated standard deviations.

\subsection{Estimating Losses}
The loss is estimated using the Beta distribution since fitting such a distribution allows the representation of risks quite accurately. The Beta distribution is a two parameter distribution, with an upper bound for the standard deviation. The standard deviation, mean, alpha and beta are defined as
\begin{eqnarray*}
\begin{array}{c c l}
\sigma_{\beta}	& = & \frac{\sigma}{max_{l}} \\
\mu_{\beta}	& = & \frac{\mu_{l}}{max_{l}} \\
\alpha 		& = & \mu_{\beta}\Big(\big({\frac{\sigma_{\beta_{max}}}{\sigma_{\beta}}\big)}^{2}-1\Big) \\
\beta 		& = & (1 - \mu_{\beta})\Big(\big({\frac{\sigma_{\beta_{max}}}{\sigma_{\beta}}\big)}^{2}-1\Big)
\end{array}
\end{eqnarray*}

An upper bound is set to limit the standard deviation using $\sigma_{\beta_{max}} = \sqrt{\mu_{\beta}(1-\mu_{\beta})}$, if $\sigma_{\beta} > \sigma_{\beta_{max}}$, then $\sigma_{\beta} = \sigma_{\beta_{max}}$. In the algorithm reported in this paper, for numerical purpose a value very close to $\sigma_{\beta_{max}}$ is chosen.

To obtain the loss after applying secondary uncertainty beta distribution functions are used as follows

\begin{eqnarray*}
\begin{array}{l c l}
Loss 				& = &	max_{l} * InvCDF_{beta}(z; \alpha, \beta)\\
InvCDF_{beta}(z; \alpha, \beta) & = &	\Big(\frac{B(z; \alpha, \beta)}{B(\alpha, \beta)}\Big)^{-1} \mbox{, where}\\
B(z; \alpha, \beta) 		& = & 	\int\limits_{0}^{z} t^{\alpha - 1}(1 - t)^{\beta - 1}\mathrm{d}t
\end{array}
\end{eqnarray*}

\section{Implementation}
\label{implementation}
In this section, the hardware platforms used for the experimental studies are firstly considered, followed by the implementation of the data structures required for aggregate risk analysis with uncertainty and the implementation of the methods for computing secondary uncertainty. Optimizations incorporated in the implementations are further considered.

Two hardware platforms are used for implementing a sequential and parallel aggregate risk analysis algorithm. Firstly, a multi-core CPU is employed whose specifications are a 3.40 GHz quad-core Intel(R) Core (TM) i7-2600 processor with 16.0 GB of RAM. The processor has 256 KB L2 cache per core, 8MB L3 cache and maximum memory bandwidth of 21 GB/sec. The processor supports hyperthreading on the physical cores making eight virtual cores available. The experiments consider virtual cores as hyperthreading is beneficial for data intensive applications. Both sequential and parallel versions of the aggregate risk analysis algorithm were implemented on this platform. The sequential version was implemented in C++, while the parallel version was implemented in C++ and OpenMP. Both versions were compiled using the GNU Compiler Collection g++ 4.7 using `\texttt{\small -O3}' and `\texttt{\small -fopenmp}' when OpenMP is used.

Secondly, an NVIDIA Tesla C2075 GPU, consisting of 448 processor cores (organized as 14 streaming multi-processors each with 32 symmetric multi-processors), each with a frequency of 1.15 GHz, a global memory of 5.375 GB and a memory bandwidth of 144 GB/sec was employed in the GPU implementations of the aggregate risk analysis algorithm. The peak double precision floating point performance is 515 Gflops whereas the peak single precision floating point performance is 1.03 Tflops. The implementation of the algorithm is compiled using the NVIDIA CUDA Compiler (nvcc), version 5.0\footnote{\small \url{https://developer.nvidia.com/cuda-toolkit}}.

The following implementations for aggregate risk analysis with uncertainty are considered in this paper: (i) a sequential implementation on the CPU, (ii) a parallel implementation on the multi-cores of the CPU, and (iii) a parallel implementation on the many-cores of the GPU. Four libraries are used for applying secondary uncertainty on the many-core GPU and four additional libraries on the multi-core CPU.

\subsection{Implementing Data Structures for the Algorithm}
In aggregate risk analysis, the losses of events in a trial need to be determined by looking up losses in the XELT. The key design question is whether the data structure containing the event-loss pairs of all trials need to be a sparse matrix in the form of a direct access table or a compact representation. While fast lookups can be obtained in the sparse matrix representation, this performance is achieved at the cost of high memory usage. Consider a YET with 1,000,000 events and one Layer with 16 XELTs, and each XELT consisting of 20,000 events with non-zero losses. The representation using a direct access table would require memory to hold $16,000,000$ event-loss pairs (without considering the data required for secondary uncertainty calculations). While such a large data structure is held in memory, 15,700,000 events represent zero loss value.

Though the sparse representation requires large amount of memory it is chosen over any compact representation for the following reason. A search operation is required to find an event-loss pair even in a compact representation. If sequential search is adopted, then $O(n)$ memory accesses are required to find an event-loss pair. If sorting is performed in a pre-processing phase to facilitate a binary search, then $O(log(n))$ memory accesses are required to find an event-loss pair. If a constant-time space-efficient hashing scheme, such as cuckoo hashing \cite{cuckoohashing} is adopted then an event-loss pair can be accessed with a constant number of memory accesses. However, this can be only be achieved at the expense of a complex implementation and overheads depreciating run-time performance. Further, such an implementation on the GPU with a complex memory hierarchy is cumbersome. Although large memory space is required for a direct access table, looking up event-loss pairs can be achieved with fewer memory accesses compared to the memory accesses in a compact representation.

Two data structure implementations of 16 XELTs were considered. In the first implementation, each XELT is considered as an independent table; therefore, in a read cycle, each thread independently looks up its events from the XELTs. All threads within a block access the same XELT. In the second implementation, all the 16 XELTs are combined into a single table. Consequently, the threads then use the shared memory to load entire rows of the combined XELTs at a time. The second implementation performs poorly compared to the first implementation. This is because of the memory overheads for the threads to collectively load rows from the combined XELT.

In the implementation on the multi-core CPU platform the entire data required for the algorithm is processed in memory. The GPU implementation of the algorithm uses the GPU's global memory to store all data structures. The parallel implementation on the GPU requires high memory transactions. This is surmounted by utilising shared memory over global memory.

\subsection{Implementing Methods to Compute Secondary Uncertainty}
Three statistical functions are required in the method for applying secondary uncertainty. They are (i) the Cumulative Distribution Function (CDF) of Normal distribution, (ii) the Quantile of the Normal distribution, and (iii) the Quantile of the Beta distribution. The Quantile of the Beta distribution is a numerically intensive function since it is an iterative method which converges to the solution within a certain error.

Seven different libraries are used for implementing the secondary uncertainty methodology on the multi-core CPU. The first is the Boost statistical library offered by the Boost C++ libraries\footnote{\small \url{http://www.boost.org/}}. The statistical functions are available inside the namespace \texttt{\small boost::math}. In order to use the distributions the header \texttt{\small <boost/math/distributions.hpp>} needs to be included. For example, \texttt{\small boost::math::normal\_distribution<> NormDist (0.0L, 1.0L)} will create a Standard Normal distribution with mean equal to 0 and standard deviation equal to 1. The Quantile function of the Normal distribution can be obtained by as \texttt{\small quantile(NormDist, double value)}. The CDF of the Normal distribution is obtained by \texttt{\small cdf (NormDist, double value)}. Similarly, an Assymetrical Beta distribution with alpha and beta values can be created using \texttt{\small boost::math::beta\_distribution<> BetaDist (double alpha, double beta)} and the Quantile can be obtained from \texttt{\small quantile(BetaDist, double cdf)}\footnote{\small \url{http://www.boost.org/doc/libs/1\_35\_0/libs/math/doc/sf\_and\_dist/html/math\_toolkit/dist/dist\_ref/dists/beta\_dist.html}}.

The second is the IMSL C/C++ Numerical Libraries offered by Rogue Wave Software\footnote{\small \url{http://www.roguewave.com/products/imsl-numerical-libraries/c-library.aspx}}. The mathematical functions are obtained from the \texttt{\small imsl.h} header file and the statistical functions are obtained from the \texttt{\small imsls.h} header file \cite{IMSL-1, IMSL-2}. The CDF for the Normal distribution with mean equal to 0 and standard deviation equal to 1 is obtained from \texttt{\small imsl\_f\_normal\_cdf (double value)} and the Quantile is obtained from \texttt{\small imsl\_f\_normal\_inverse\_cdf (double cdf)}. The Quantile for the Beta distribution with alpha and beta values are obtained as \texttt{\small imsl\_f\_beta\_inverse\_cdf (double cdf, double alpha, double beta)}.

The third is PROB which is a C++ library that handles the Probability Density Functions for various discrete and continuous distributions\footnote{\small \url{http://people.sc.fsu.edu/~jburkardt/cpp\_src/prob/prob.html}}. In order to use the distributions the header \texttt{\small prob.hpp>} needs to be included. The CDF for the Normal distribution with mean equal to 0 and standard deviation equal to 1 is obtained from \texttt{\small normal\_01\_cdf (double value)} or \texttt{\small normal\_cdf (double x, double a, double b)} (where $a = 0$ and $b = 1$) and the Quantile from \texttt{\small normal\_01\_cdf\_inv (double cdf)} or \texttt{\small normal\_cdf\_inv (double cdf, double a, double b)} (where $a = 0$ and $b = 1$). The Quantile for the Beta distribution with alpha and beta values are obtained as \texttt{\small beta\_cdf\_inv (double cdf, double alpha, double beta)} \cite{ASA109}.

The fourth is DCDFLIB which is a C library adapted from Fortran for evaluating CDF and inverse CDF of discrete and continuous probability distributions\footnote{\small \url{http://www.netlib.org/random/}}. In order to use the distributions the header \texttt{\small dcdflib.c} needs to be included. The CDF and the Quantile for the Normal distribution with mean stored in \texttt{\small mean} and standard deviation stored in \texttt{\small sd} can be obtained from \texttt{\small cdfnor (int *which, double *p, double *q, double *value, double *mean, double *sd, int *status, double *bound)} \cite{DCDFLIB-2}. `\texttt{\small which}' is set to 1 to obtain the CDF value \texttt{\small p} and \texttt{\small q = 1.0 - p}. `\texttt{\small which}' is set to 2 to obtain the Quantile in \texttt{\small value}. \texttt{\small status} and \texttt{\small bound} are variables to report the status of the computation. The Quantile of the Beta distribution \texttt{\small x} and \texttt{\small y = 1.0 - x} for \texttt{\small alpha} and \texttt{\small beta} can be obtained from \texttt{\small cdfbet (int *which, double *p, double *q, double *x, double *y, double *alpha, double *beta, int *status, double *bound)} when \texttt{\small which} is set to 2, \texttt{\small p} is the CDF and \texttt{\small q = 1.0 - p} \cite{DCDFLIB-3}.

The fifth library is ASA310 or the Applied Statistics Algorithm 310\footnote{\small \url{http://lib.stat.cmu.edu/apstat/}} which is a C++ library for evaluating the CDF of the Noncentral Beta distribution \cite{ASA310-1}. The include file is \texttt{\small asa310.hpp}. The iterative algorithm for achieving convergence of the solution to compute the Quantile calls the function for computing the tail of the Noncentral Beta distribution \texttt{\small betanc (float value, float alpha, float beta, float\\lambda, int *ifault)}, where \texttt{\small lambda}, the noncentrality parameter is set to 0 for the standard Beta distribution and \texttt{\small ifault} is an error flag.

The sixth library is ASA226 or the Applied Statistics Algorithm 226\textsuperscript{7} which is a C++ library similar to ASA310 \cite{ASA226-1, ASA226-2}. The include file is \texttt{\small asa226.hpp}. The iterative algorithm for achieving convergence of the solution to compute the Quantile is used to call the function for computing the tail of the Noncentral Beta distribution.

The seventh library is BETA\_NC another C++ library that can evaluate the CDF of the Noncentral Beta distribution \cite{betanc}. The include file is \texttt{\small beta\_nc.cpp}. The iterative algorithm that achieves convergence of the solution to compute the Quantile calls the \texttt{\small beta\_noncentral\_cdf (double alpha, double beta, double lambda, double value, double error\_max)}, where \texttt{\small lambda}, the noncentrality parameter is set to 0 for the standard Beta distribution and \texttt{\small error\_max} is is the error control in the computation.

For implementing the secondary uncertainty methodology on the GPU statistical functions provided by the CUDA Math API are employed by including the \texttt{\small math.h} header file\footnote{\small \url{http://docs.nvidia.com/cuda/cuda-math-api/index.html}}. The CDF of the Normal distribution \texttt{\small normcdf} and the Quantile of the Normal Distribution \texttt{\small normcdfinv} are fast methods and included in the implementation. The CUDA Math API currently does not support Beta distribution functions. Therefore, four libraries, namely the PROB, ASA310, ASA226 and BETA\_NC are incorporated in the implementation for the many-core GPU. These libraries are ported for the GPU platform and all the functions in the libraries are implemented as \texttt{\small \_\_device\_\_} functions for the GPU.

\begin{figure}
	\centering
	\includegraphics[width = 0.48\textwidth]{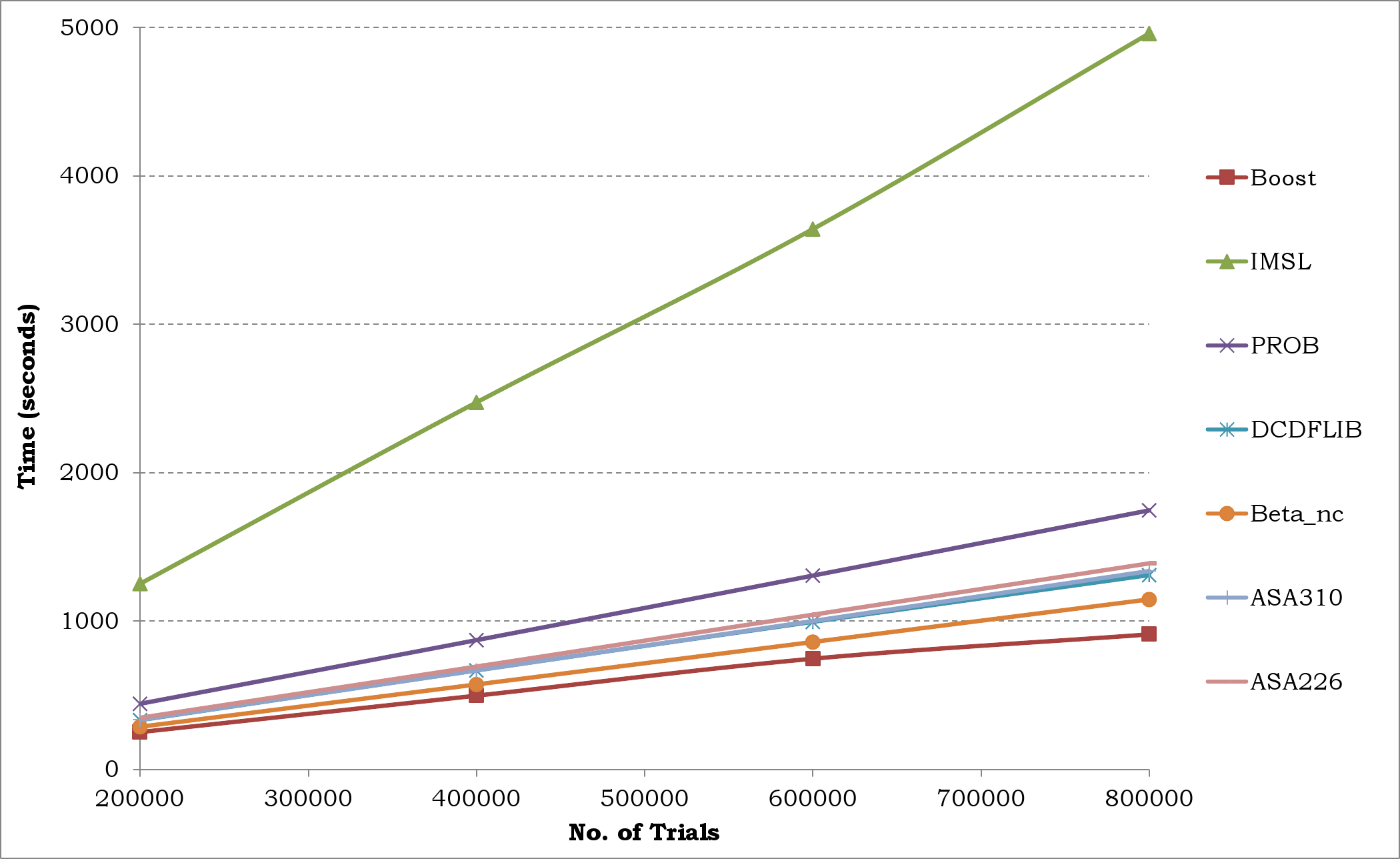}
	\caption{Applying secondary uncertainty using different statistical libraries for sequential implementation on CPU}
	\label{figure1}
\end{figure}

\subsection{Optimising the Implementation}
The implementations were optimised for better performance in three ways. Firstly, by incorporating loop unrolling, which refers to the compiler replicating of blocks of code within `for loops' to reduce the number of iterations performed by for loops. The \texttt{\small for} loops are unrolled using the \texttt{\small pragma} directive; the \texttt{\small for} loops in line nos. 1-5 of Algorithm \ref{algorithm1} can be unrolled as each iteration is a mutually independent iteration.

Secondly, in the case of the GPU by migrating data from both shared and global memory to the kernel registry. The kernel registry has the lowest latency compared to all other memory.

Thirdly, by reducing the precision of variables used in the algorithm, whereby the double variables are changed to float variables. In the case of the GPU, read operations are faster using float variables as they are only half the size of a double variable. Furthermore, the performance of single precision operations tend to be approximately twice as fast as double precision operations. The CUDA Math API supports functions for floating point operations and the full acceleration of CUDA Math API can be achieved by using the compiler flag \texttt{\small -use\_fast\_math}.

\section{Experimental Results}
\label{results}
In this section, the results obtained from the sequential implementation on the CPU, the parallel implementation on the multi-core CPU and the many-core GPU, and the summary of the experimental results are presented. 

Figure \ref{figure1} shows the graph plotted for the time taken for sequentially performing aggregate risk analysis using trials varying from 200,000 to 800,000 with each trial comprising 1,000 events on the CPU when secondary uncertainty is applied using the Boost, IMSL, PROB, DCDFLIB, BETA\_NC, ASA310 and ASA226 libraries. The experiments are performed for one Layer and 16 XELTs. The Boost library provides the fastest functions for secondary uncertainty followed by the BETA\_NC, DCDFLIB, ASA310 and ASA226 libraries. The PROB library is approximately 2 times slower and the IMSL numerical library is approximately 5 times slower than the Boost library.

\begin{figure}
	\centering
	\includegraphics[width = 0.48\textwidth]{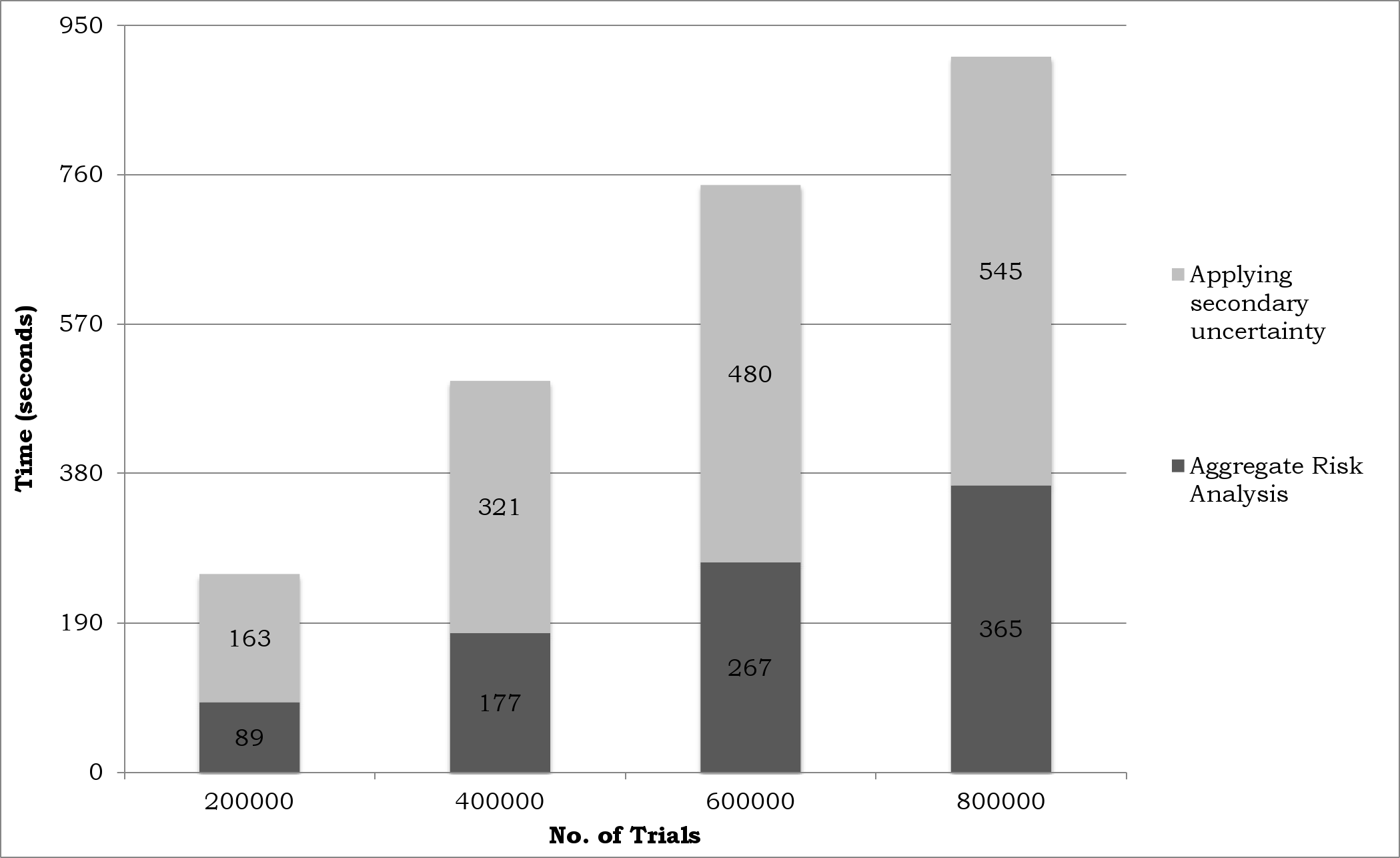}
	\caption{Time taken for aggregate risk analysis and applying secondary uncertainty for different trials in sequential implementation on CPU}
	\label{figure2}
\end{figure}

Figure \ref{figure2} shows the graph plotted for the time taken for applying secondary uncertainty for trials varying from 200,000 to 800,000 with each trial comprising 1,000 events on the CPU when Boost library is used. The results from the Boost library are chosen since it provides the fastest functions for applying secondary uncertainty. The experiments are performed for one Layer and 16 XELTs. In each case of trials shown in the graph the time for applying secondary uncertainty is nearly 2.5 times the time taken for aggregate risk analysis. The mathematical functions employed for applying secondary uncertainty are fast methods with the exception of the inverse cumulative distribution function of the beta distribution which takes majority of the time.

The time taken both for performing aggregate risk analysis with only primary uncertainty and for applying secondary uncertainty with increasing number of trials should scale linearly and this is observed both in Figure \ref{figure1} and Figure \ref{figure2}.

\begin{figure}
	\centering
	\includegraphics[width = 0.48\textwidth]{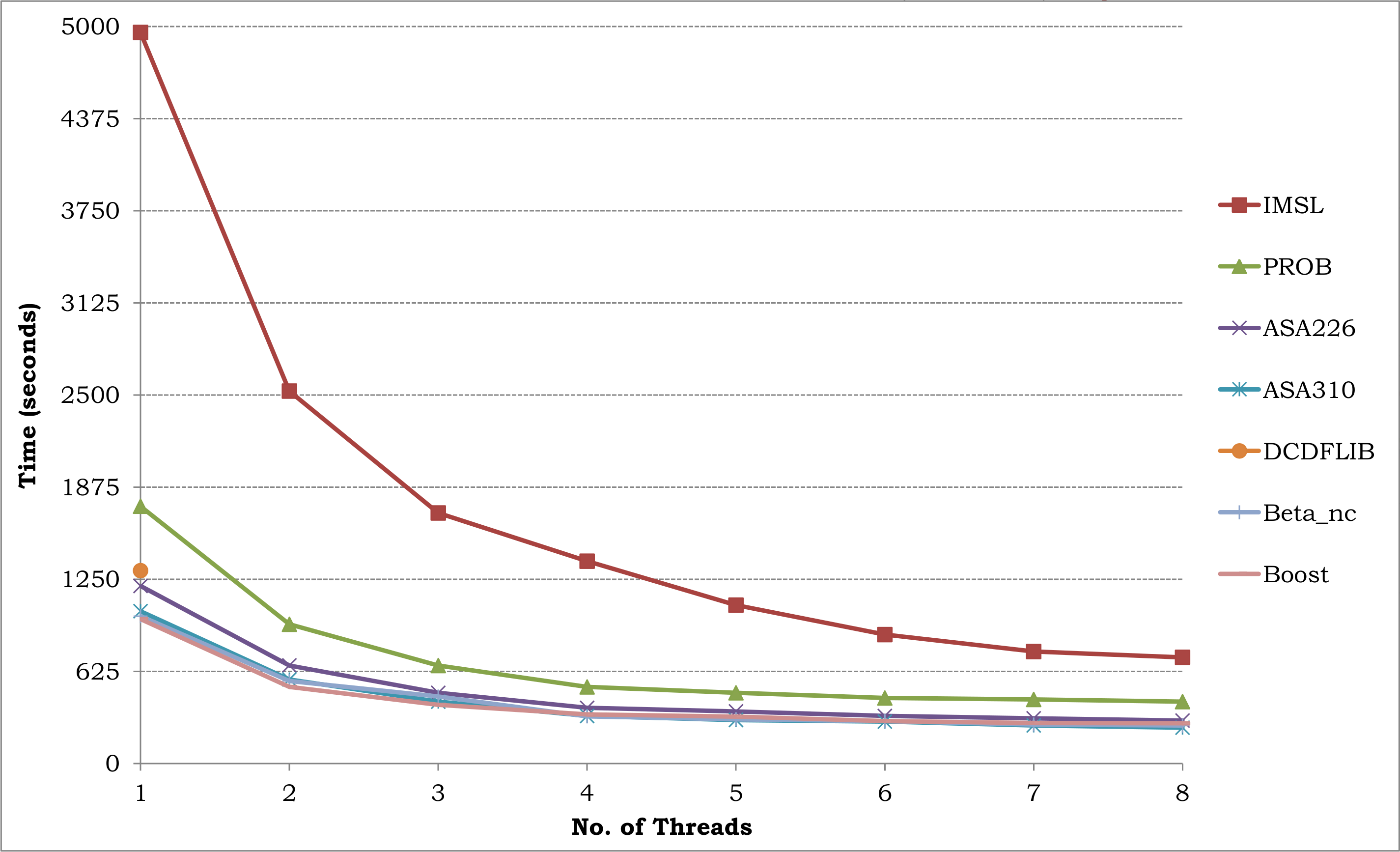}
	\caption{Time taken for aggregate risk analysis and applying secondary uncertainty for one thread on each virtual core of the CPU}
	\label{figure3}
\end{figure}

\begin{figure}
	\centering
	\includegraphics[width = 0.48\textwidth]{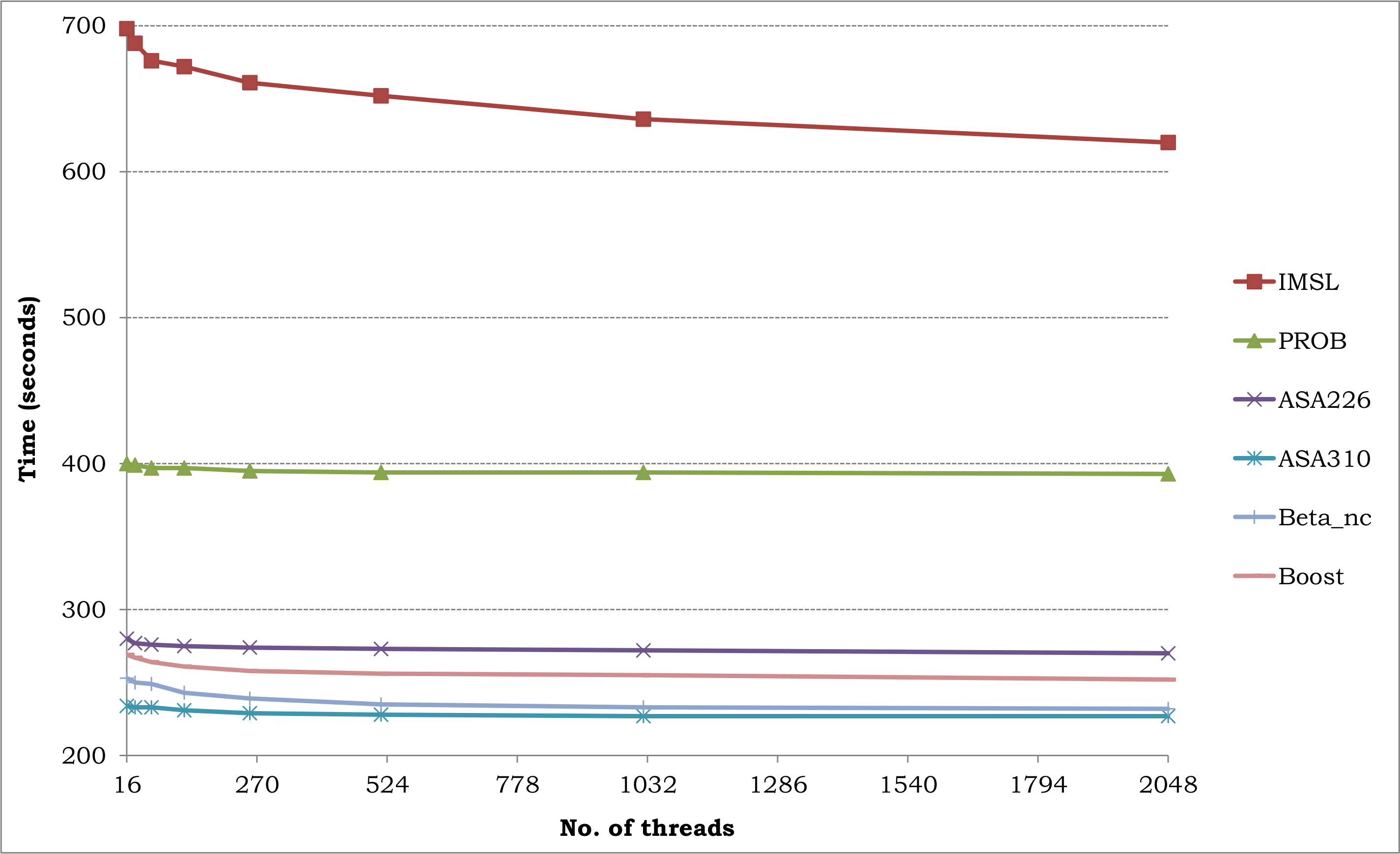}
	\caption{Time taken for aggregate risk analysis and applying secondary uncertainty for multiple threads on each virtual core of the CPU}
	\label{figure4}
\end{figure}

Figure \ref{figure3} and Figure \ref{figure4} show the graphs plotted for the time taken for performing parallel aggregate risk analysis and applying secondary uncertainty for 800,000 trials on the multi-core CPU using the Boost, IMSL, PROB, DCDFLIB, BETA\_NC, ASA310 and ASA226 libraries.

In Figure \ref{figure3}, a single thread is run on each virtual core of the CPU and the number of cores are varied from 1 to 8 (i.e., up to two threads on each of the four physical cores). Each threads performs the aggregate risk analysis for a single trial and applies secondary uncertainty. Multiple threads are used by employing OpenMP directive \texttt{\small \#pragma omp parallel} in the C++ source. With respect to the overall time the ASA310 library performs the best requiring 232 seconds for the analysis. For the IMSL library, a speedup of nearly 1.9x is achieved for two cores, a speedup of nearly 3.6x is obtained for four cores and a speedup of 6.9x is obtained for 8 cores. While the performance keeps diminishing for the IMSL library, the efficiency of all the other libraries used in secondary uncertainty is significantly low. No more than 4x speedup is achieved on eight cores in the best case. The limiting factor is that the bandwidth to memory is not increased as the number of cores increase. The majority of the time taken in aggregate risk analysis is for performing random access reads into the data structure representing the XELT. The majority of the time in applying secondary uncertainty is consumed in the inverse cumulative distribution function of the beta distribution. The Boost library outperforms all the other libraries with respect to the overall time. The DCDFLIB library did not scale on multiple threads as the files are written as blocks of program with unconditional jumps using the \texttt{\small goto} statement.

In Figure \ref{figure4}, the performance on all eight virtual cores of the CPU is illustrated; multiple threads are run on each virtual core of the CPU. For example, when 16 threads are employed two threads run on each virtual core and when 2048 threads are employed 256 threads run on each virtual core of the CPU. A small drop is observed in the absolute time when many threads are executed on each core. When 256 threads run on a core, the overall runtime drops from 701 seconds (using two threads per core) to 620 seconds for the IMSL library and the runtime drops from 269 seconds (using two threads per core) to 252 seconds for the Boost library. When 2048 threads are employed, the ASA310 library performs better than the Boost library by 25 seconds.

\begin{figure}
	\centering
	\includegraphics[width = 0.48\textwidth]{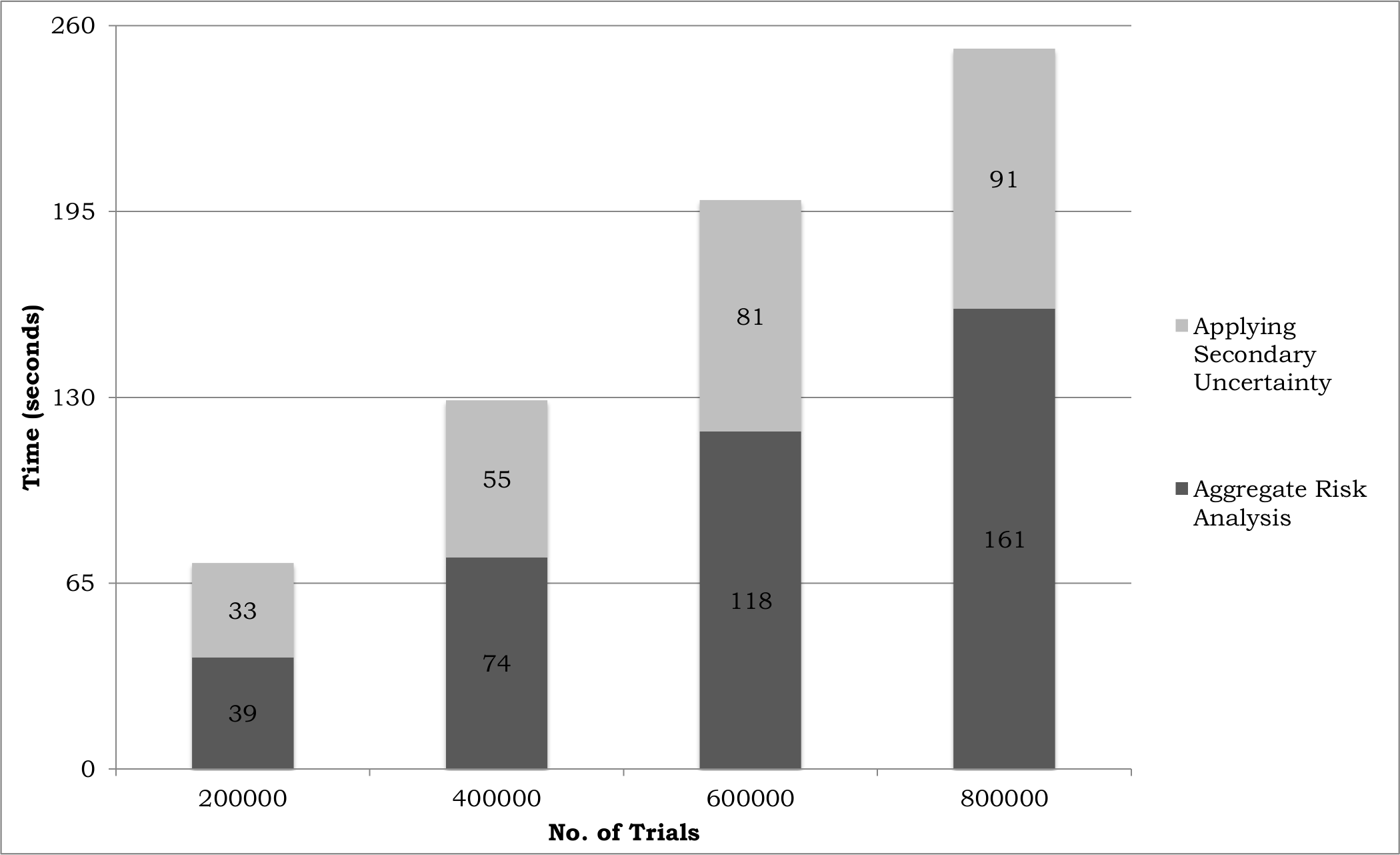}
	\caption{Time taken for aggregate risk analysis and applying secondary uncertainty for different trials in parallel implementation on multi-core CPU}
	\label{figure5}
\end{figure}

Figure \ref{figure5} shows the graph plotted for the time taken for applying secondary uncertainty using 2048 threads for trials varying from 200,000 to 800,000 on the eight virtual cores of the CPU when the Boost library is used. In each case of trials shown in the graph the time taken for applying secondary uncertainty increases with the number of trials. The time taken for applying secondary uncertainty on the multi-core is only $\frac{1}{6}^{th}$ the time taken in the sequential implementation.

\begin{figure}
	\centering
	\includegraphics[width = 0.48\textwidth]{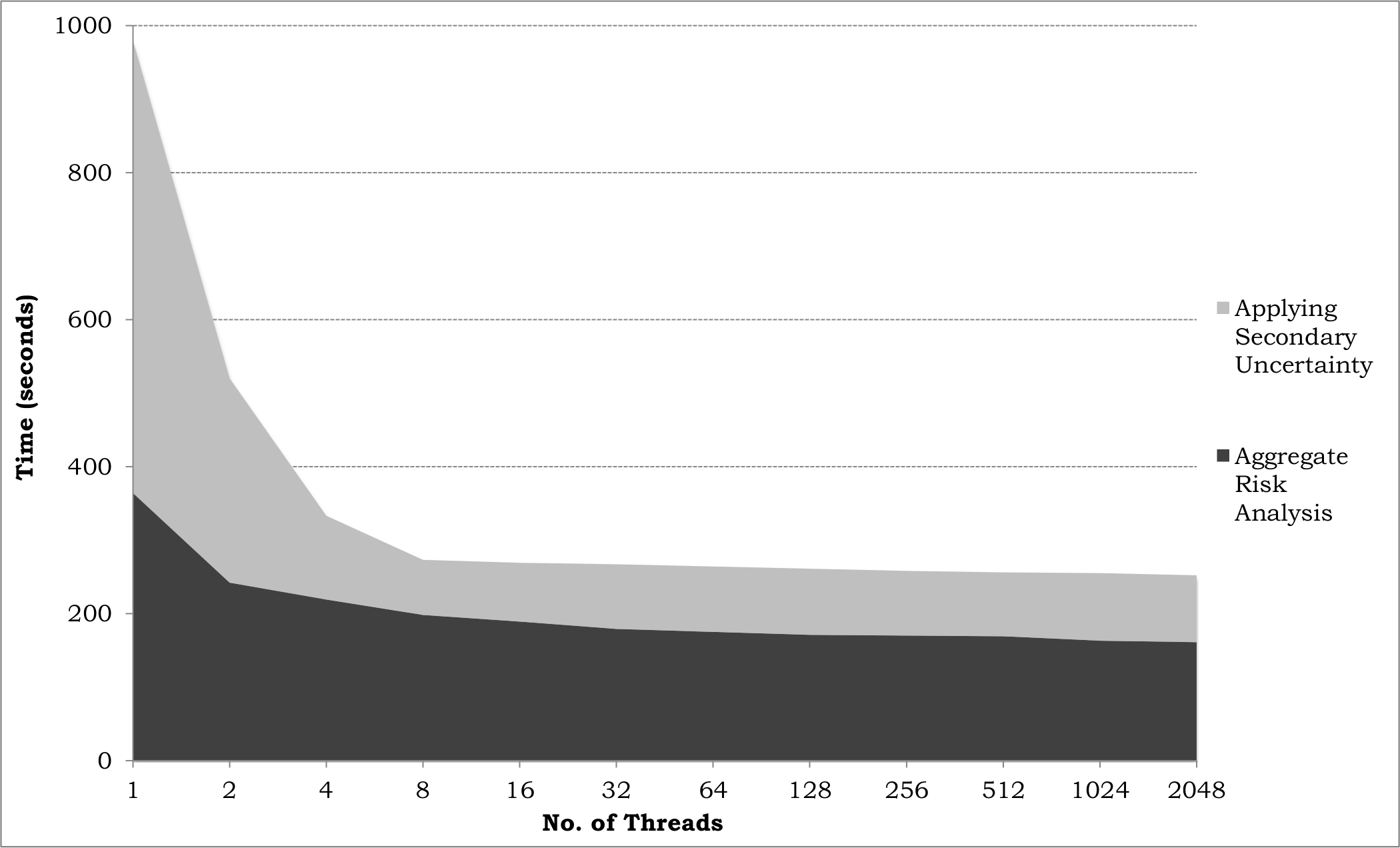}
	\caption{Time taken for aggregate risk analysis and applying secondary uncertainty using Boost library for 800,000 trials in parallel implementation on multi-core CPU}
	\label{figure6}
\end{figure}

Figure \ref{figure6} shows the graph plotted for the time taken for aggregate risk analysis and for applying secondary uncertainty using the Boost library for 800,000 trials when the number of threads are varied from 1 to 2048 on the multi-core CPU. The lowest overall time is 252 seconds when 256 threads are employed per core of the CPU; 161 seconds are required for the aggregate risk analysis and 91 seconds for applying secondary uncertainty. However, the lowest time taken for applying secondary uncertainty is 75 seconds which is achieved when one thread is used per core (8 threads on the CPU). While there is a decrease in the overall time taken as the number of threads increase, there is a gradual increase in the time taken for applying secondary uncertainty when more than 8 threads are employed on the CPU; 75 seconds when 8 threads are used, where as 91 seconds required when 2048 threads are used. This is due to the increasing overhead in swapping constants in and out of memory as the number of threads increase. The performance of Boost surpasses that of ASA only when one and two threads are used. Beyond two threads ASA310 has lower overall time.

\begin{figure}
	\centering
	\includegraphics[width = 0.48\textwidth]{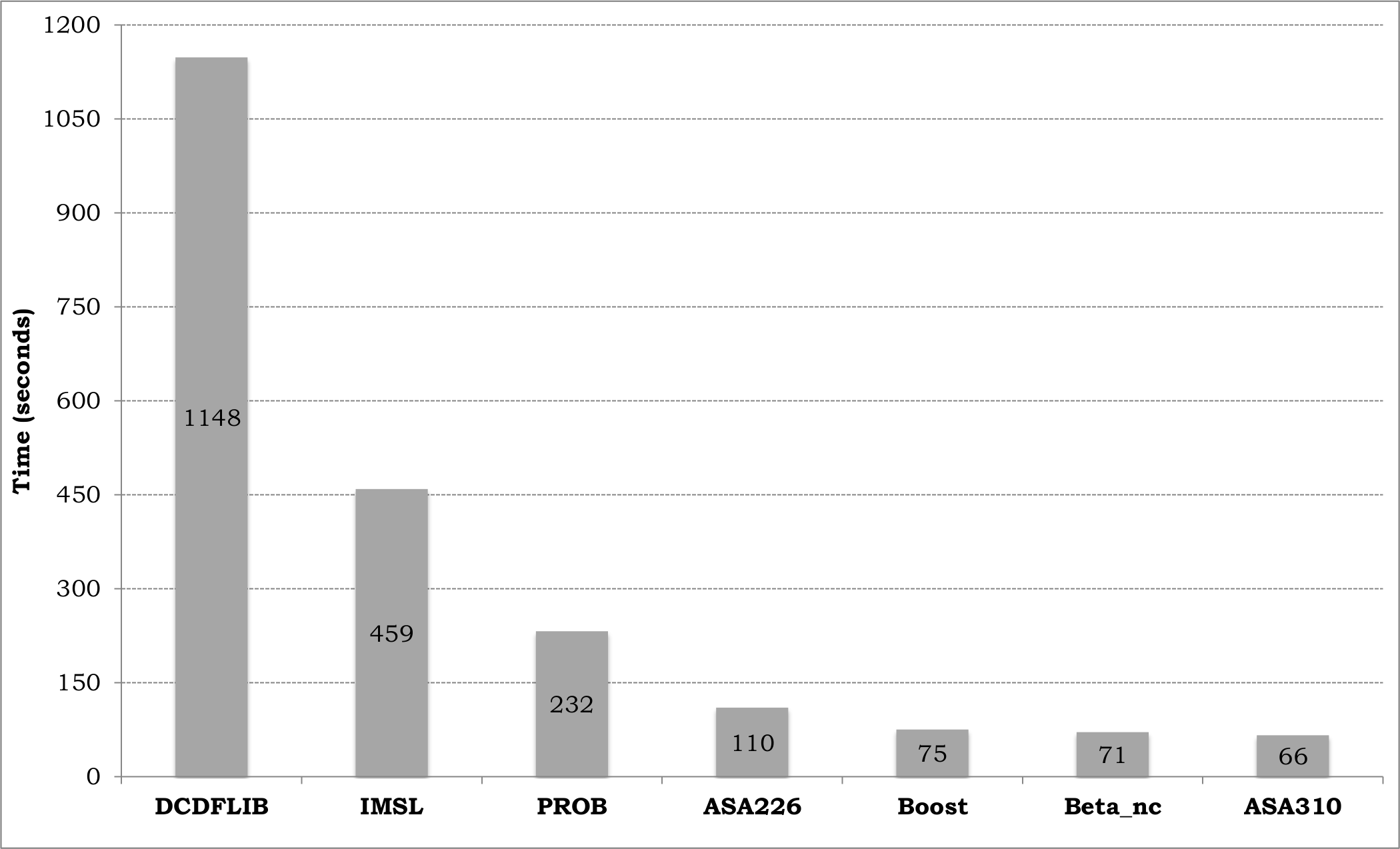}
	\caption{Optimal time taken for applying secondary uncertainty using different libraries for 800,000 trials in parallel implementation on multi-core CPU}
	\label{figure7}
\end{figure}

Figure \ref{figure7} shows the graph plotted for the optimal (best time) time taken for applying secondary analysis in aggregate risk analysis for 800,000 trials using different libraries. Optimality for overall time is achieved when 2048 threads are employed; in this graph the optimality for applying secondary uncertainty is considered. For the ASA310 library, the best time for applying secondary uncertainty is 66 seconds which is only $\frac{1}{8}^{th}$ the time taken in the sequential implementation.

\begin{figure}
	\centering
	\includegraphics[width = 0.48\textwidth]{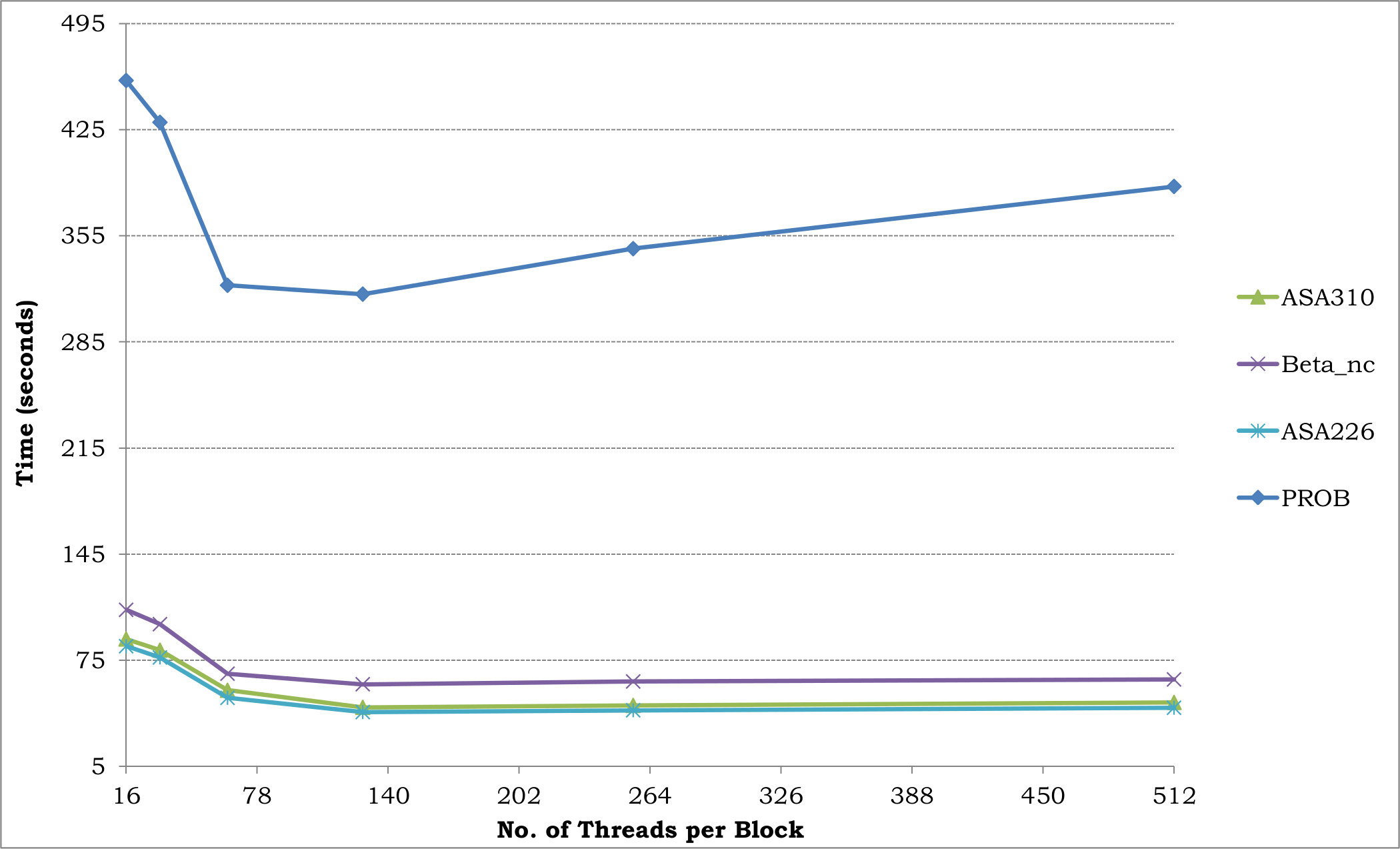}
	\caption{Time taken for aggregate risk analysis and applying secondary uncertainty using different threads per block on many-core GPU}
	\label{figure9}
\end{figure}

\begin{figure}
	\centering
	\includegraphics[width = 0.48\textwidth]{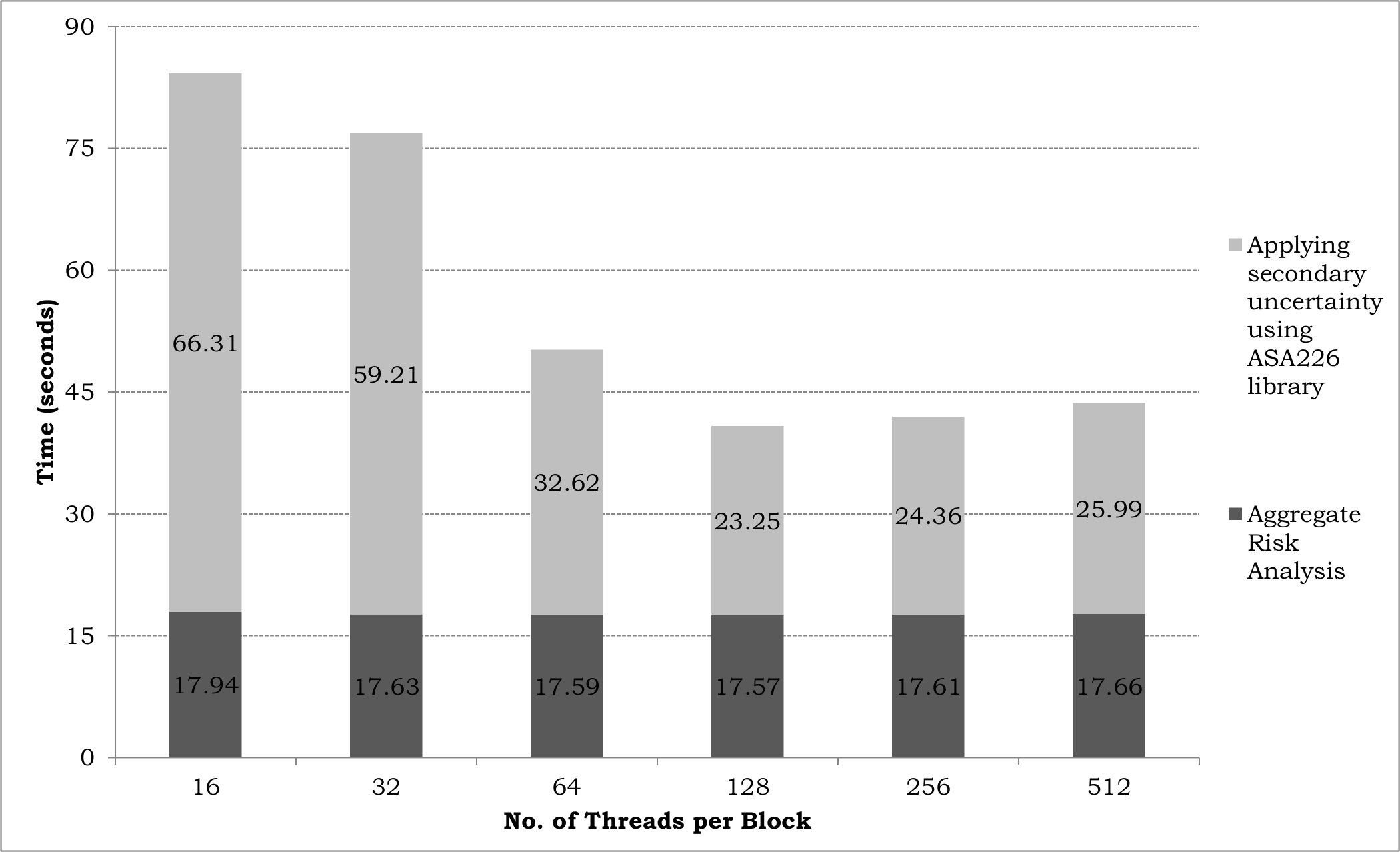}
	\caption{Time taken for aggregate risk analysis and for applying secondary uncertainty using different threads per block for ASA226 on many-core GPU}
	\label{figure10}
\end{figure}

Figure \ref{figure9} and Figure \ref{figure10} show the graphs plotted for the time taken for performing parallel aggregate risk analysis and applying secondary uncertainty for 800,000 trials on the many-core GPU using the PROB, ASA310, BETA\_NC and ASA226 libraries. IMSL and Boost libraries are not available for GPUs. The DCDFLIB library was ported for the GPU but did not execute on the hardware.

CUDA provides abstraction over the streaming multi-processors of the GPU, which is often referred to as a CUDA block. The number of threads executed per CUDA block can be varied in aggregate risk analysis. For example, consider the execution of 800,000 Trials using 800,000 threads. If 256 threads are executed on one Streaming Multi-Processor (SMP), then 3125 CUDA blocks need to be executed on the 14 SMPs. Each SMP will have to execute 223 CUDA blocks. All threads executing on one SMP have a fixed size of shared and constant memory. Fewer the threads employed, then each thread will have a large size of shared and constant memory. But there is a trade-off when fewer threads are used since the latency for accessing the global memory of the GPU increases.

In Figure \ref{figure9} the analysis is performed for 800,000 Trials on the GPU by varying the number of threads per block from 16 to 512 threads. An improvement in the performance is seen as the number of threads increase from 16 to 128 since the latency for accessing the global memory drops. Beyond 128 threads the performance starts to drop as the shared and constant memory available to each thread decreases. ASA226 library performs the best since the function used in computing secondary uncertainty has an optimal balance between the number of constants and the amount of computation required. This is vital when there is a trade-off between the size of shared and constant memory and latency in accessing global memory. The lowest time taken is 51 seconds when 128 threads per block are used.

Figure \ref{figure10} shows the time taken to perform aggregate risk analysis for applying primary uncertainty and for applying secondary uncertainty using the ASA226 library on the GPU for 800,000 Trials. The time taken for performing aggregate risk analysis is nearly a constant. The time taken for applying secondary uncertainty first decreases from 16 to 128 threads per block and then increase beyond 128 threads per block. This is due to the trade-off between the size of the shared and constant memory and latency in accessing global memory. In the best case when 128 threads per block are employed the time taken for applying secondary uncertainty is nearly twice the time taken for performing aggregate risk analysis. The best time for applying secondary uncertainty using ASA226 on the GPU is only half the best time taken by ASA310 for applying secondary uncertainty using multiple threads on the CPU.

\begin{figure}
	\centering
	\includegraphics[width = 0.48\textwidth]{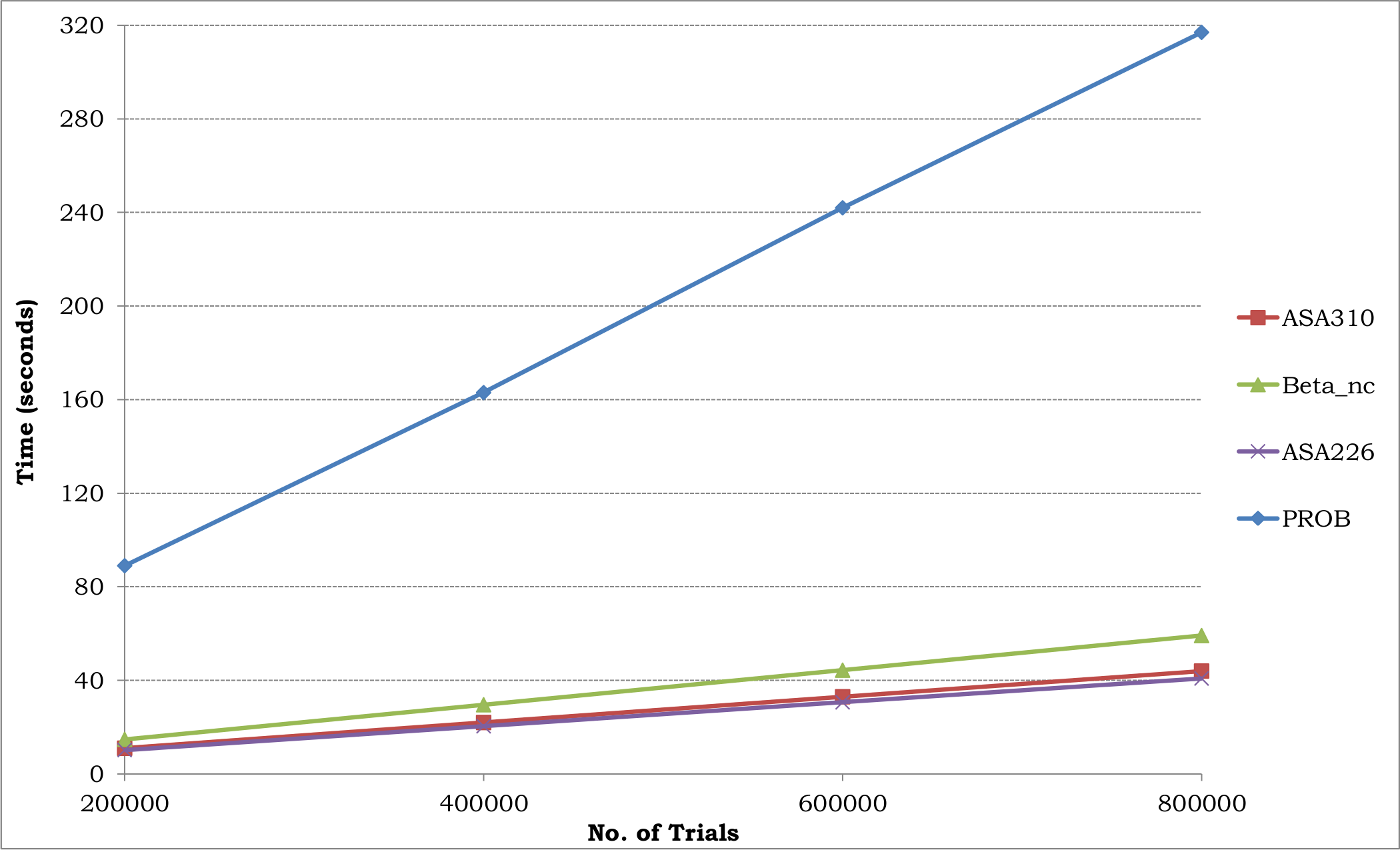}
	\caption{Time taken for aggregate risk analysis and applying secondary uncertainty for different trials in parallel implementation on many-core GPU}
	\label{figure8}
\end{figure}

\begin{figure}
	\centering
	\includegraphics[width = 0.48\textwidth]{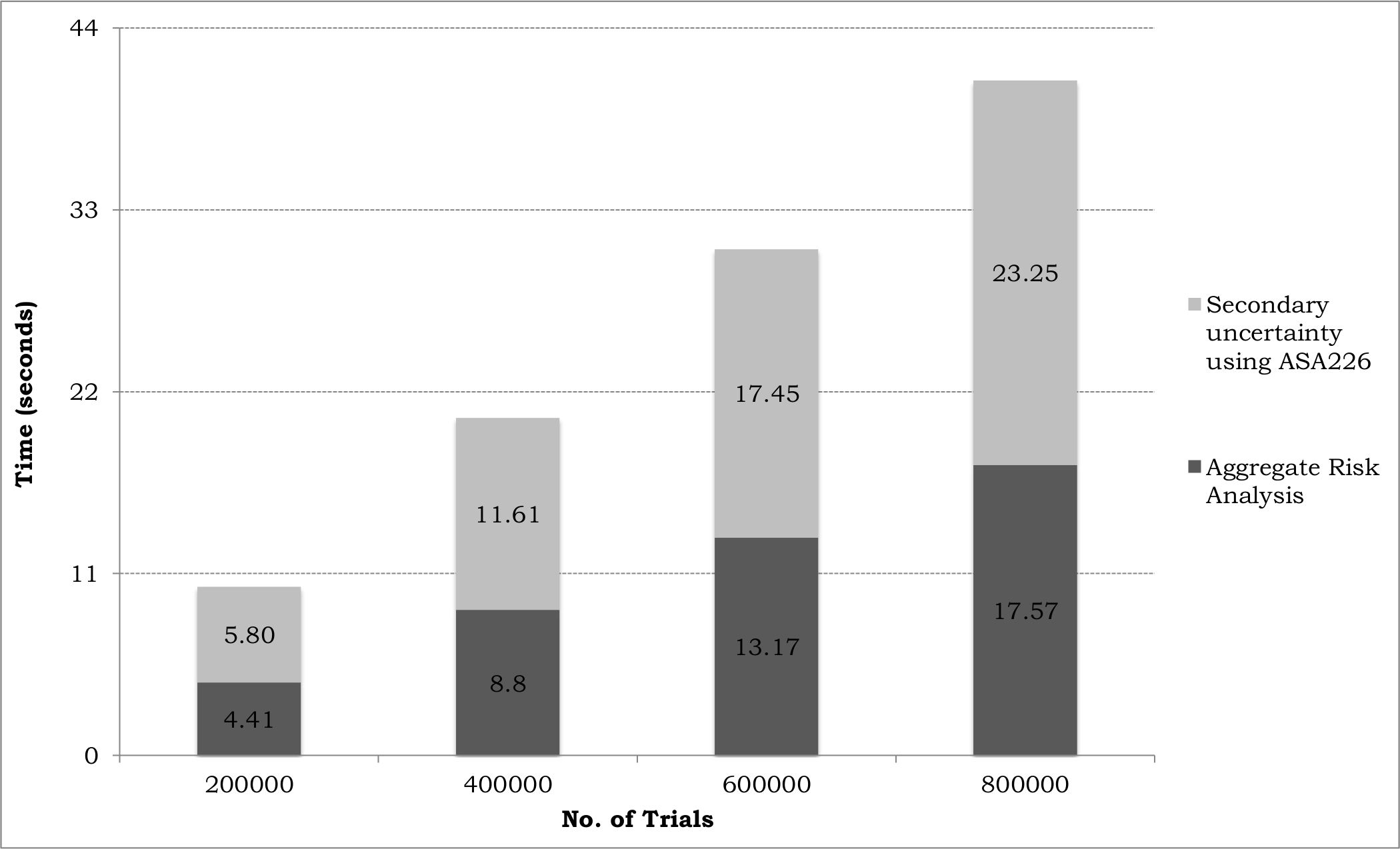}
	\caption{Time taken for aggregate risk analysis and for applying secondary uncertainty for different trials using ASA226 on many-core GPU}
	\label{figure11}
\end{figure}

Figure \ref{figure8} illustrates the performance of the ASA226, BETA\_NC, ASA310 and PROB libraries on the GPU for different trials varying from 200,000 to 800,000. In each case of trials the time taken for applying secondary uncertainty increases with the number of trials. The ASA226 outperforms the BETA\_NC, ASA310 and PROB libraries. Figure \ref{figure11} shows the time taken to perform aggregate risk analysis for applying primary uncertainty and for applying secondary uncertainty using the ASA226 library on the GPU for trials varying from 200,000 to 800,000. Both times scale linearly.

\subsection{Discussion}
\begin{figure}
	\centering
	\includegraphics[width = 0.48\textwidth]{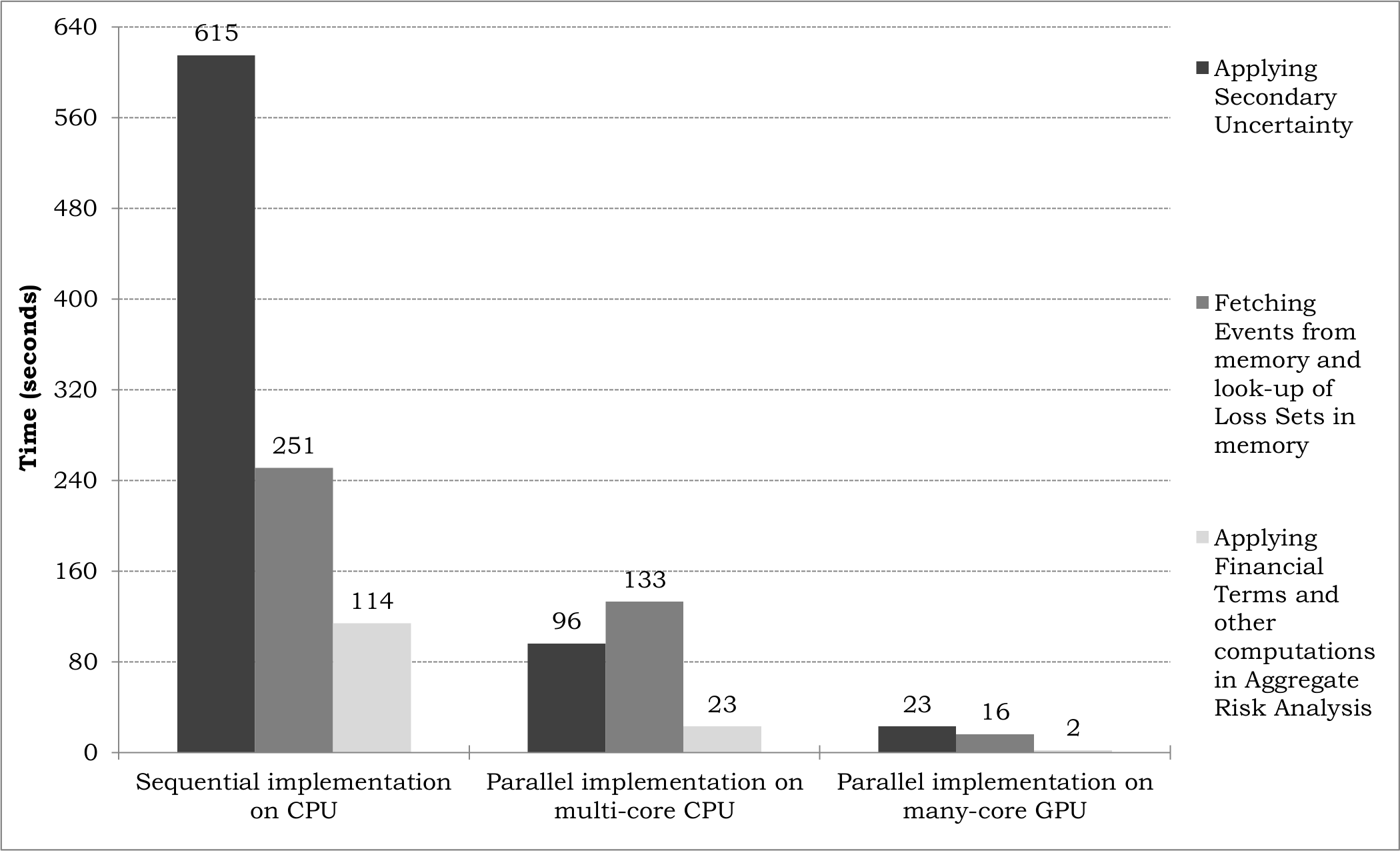}
	\caption{Time taken for fetching Events from memory and for look up of Loss Sets in memory, applying Financial Terms and performing other computations in aggregate risk analysis, and applying secondary uncertainty on sequential implementation on CPU and parallel implementations on CPU and GPU}
	\label{figure12}
\end{figure}

Figure \ref{figure12} is a graph that summarises the key results from the experimental study. The set of three bars represents the time taken for (i) fetching Events from memory and for look up of Loss Sets in memory, (ii) applying Financial Terms and performing other computations in aggregate risk analysis, and (iii) applying secondary uncertainty on the sequential implementation on the CPU and the parallel implementations on both the multi-core CPU and many-core GPU when 800,00 Trials, with each Trial consisting 1,000 Events, and 16 XELTs are employed. In each case, parameters specific to the implementation, such as the number of threads, were set to the best value identified during experimentation.

In the parallel implementations for the basic aggregate analysis, a speedup of 2.3x is achieved on the CPU and a speedup of 20x is achieved on the GPU when compared against the sequential implementation. A speedup of 24x is achieved in the overall time for the implementation on the GPU in contrast to the sequential implementation on the CPU. For applying secondary uncertainty, multiple threading on the eight virtual cores of the CPU is nearly five times faster than the sequential implementation and three times slower than the GPU. For the numeric computations on the GPU an accelaration of approximately 26x is achieved over the sequential implementation. Limited memory bandwidth is a bottleneck in the CPU resulting in approximately 27\% and 53\% of the time being spent for fetching Events and for look up of Loss Sets in memory for the sequential and parallel implementation on the CPU respectively. While the time for fetching Events and for look up of Loss Sets in memory have been significantly lowered on the GPU 39\% of the total time is still used to this end.

In the sequential implementation on the CPU, in the parallel implementation on the multi-core CPU and in the parallel implementation on the many-core GPU approximately 62\%, 38\% and 56\%, respectively, of the total time for aggregate risk analysis is required for applying secondary uncertainty. The majority of this time is required by the computations of the Inverse Beta Cumulative Distribution. This calls for not only the development of fast methods to apply secondary uncertainty in risk analytics, but also the development of fast methods for the underlying statistical functions. Fast methods have been implemented for computing the inverse CDF of the symmetrical beta distribution \cite{invsymbetadist} which considers one shape parameter, but there are minimal implementations of fast assymetrical beta distribution that takes two shape parameters.

\section{Conclusions}
\label{conclusion}

The research reported in this paper was motivated towards experimentally verifying whether GPUs can accelerate aggregate risk analysis with both primary and secondary uncertainty. To this end an algorithm for the analysis of portfolios of risk and a methodology for applying secondary uncertainty was proposed and implemented. A sequential and a parallel implementation on the CPU and a parallel implementation on the GPU were presented. Seven statistical libraries, namely Boost, IMSL, DCDFLIB, PROB, ASA310, ASA226 and BETA\_NC were investigated for implementing the computations of secondary uncertainty. Numerous challenges in handling large data in limited memory of the GPU were surmounted; the resultant was a speedup of 24x which was achieved for the parallel analysis on the GPU over its sequential counterpart on the CPU.

In this research, the GPU performed well for the numerical computations of secondary uncertainty; the GPU was five times faster than the multiple threaded analysis on the multi-core CPU for applying secondary uncertainty. The CPU could have performed well had it not been for its limited memory bandwidth and the GPU could have performed better had it not been for its limited memory availability. 

\bibliographystyle{abbrv}

\end{document}